\pdfoutput=1 % for arXiv to use pdflatex
\documentclass[twocolumn,epjc3]{svjour3}
\RequirePackage[T1]{fontenc}

\smartqed  % flush right qed marks, e.g. at end of proof

\RequirePackage{graphicx}
\RequirePackage{mathptmx}      % use Times fonts if available on your TeX system
\RequirePackage{flushend}
\RequirePackage[numbers,sort&compress]{natbib}

\journalname{Eur. Phys. J. C}
\usepackage{amsmath,amssymb}
\usepackage{epsfig,color}
\usepackage{enumerate}
\usepackage{bm}
\usepackage{dcolumn}
\usepackage{url}
\usepackage{multirow}
\usepackage{rotating}
\usepackage{heppennames2}
\usepackage{units}
\usepackage{epstopdf}
\RequirePackage[colorlinks,citecolor=blue,urlcolor=blue,linkcolor=blue]{hyperref}

\newcolumntype{.}{D{.}{.}{1}}
\graphicspath{ {./Figures/} }
\newcommand{\ee}{\ensuremath{\Pep\Pem}\xspace}%
\newcommand{\ttZ}{\ensuremath{\PQt\PAQt\PZ}\xspace}%
\newcommand{\ttbar}{\ensuremath{\PQt\PAQt}\xspace}
\newcommand{\bbbar}{\ensuremath{\PQb\PAQb}\xspace}
\newcommand{\ytop}{\ensuremath{y_\mathrm{t}}}

\newcommand{\abinv}{\ensuremath{\mathrm{ab}^{-1}}\xspace}
\newcommand{\smH}{\PH\xspace}
\newcommand{\ttH}{\ensuremath{\PQt\PAQt\smH}\xspace}
\newcommand{\toppair}{\ensuremath{\PQt\PAQt}\xspace}
\newcommand{\bpair}{\ensuremath{\PQb\PAQb}\xspace}
\newcommand{\gghadrons}{\mbox{\ensuremath{\PGg\PGg \rightarrow \mathrm{hadrons}}}\xspace}
\newcommand{\epluseminus}{\ensuremath{\Pep\Pem}\xspace}
\newcommand{\micron}{\ensuremath{\upmu\mathrm{m}}}

\begin{document}
%
% Title -- need texorpdfstring to appease the hyperref package (math symbols in title or heading)
%
\title{Full simulation study of the top Yukawa coupling at the ILC at \texorpdfstring{$\sqrt{s}=1$~TeV}{√s=1~TeV}}
\author{T. Price\thanksref{e1,addr1}
    \and
        P. Roloff\thanksref{e2,addr2}
    \and
        J. Strube\thanksref{e3,addr3,addr5}
    \and
        T. Tanabe\thanksref{e4,addr4}
    }
\thankstext{e1}{e-mail: t.price@bham.ac.uk}
\thankstext{e2}{e-mail: philipp.roloff@cern.ch}
\thankstext{e3}{e-mail: jan.strube@pnnl.gov}
\thankstext{e4}{e-mail: tomohiko@icepp.s.u-tokyo.ac.jp}

\institute{University of Birmingham, United Kingdom\label{addr1}
          \and
          CERN, CH-1211 Geneva 23, Switzerland\label{addr2}
          \and
          Tohoku University, Aramaki, Aoba-ku, Sendai 980-8578, Japan\label{addr3}
          \and
          ICEPP, The University of Tokyo, Bunkyo-ku, Tokyo 113-0033, Japan\label{addr4}
          \and
          Pacific Northwest National Laboratory, PO Box 999, Richland, WA, 99352\label{addr5}
}

\date{Received: date / Accepted: date}
% The correct dates will be entered by the editor
\maketitle
%
% Abstract
%
\begin{abstract}
We present a study of the expected precision for the measurement of the top Yukawa coupling, $\ytop$,
in $\epluseminus$ collisions at a center-of-mass energy of \unit[1]{TeV}.
% and assuming a beam polarization of $P(\Pem,\Pep)=(-0.8,+0.2)$.
Independent analyses of $\ttH$ final states containing at least six hadronic jets are performed,
based on detailed simulations of SiD and ILD, the two candidate detector concepts for the ILC.
We estimate that a statistical precision on $\ytop$ of $4.5\%$ can be obtained with an
integrated luminosity of \unit[1]{\abinv} that is split equally between two configurations for the beam polarization $P(\Pem,\Pep)$,
 $(-80\%,+20\%)$ and $(+80\%,-20\%)$.
This estimate improves to $4\%$ if the \unit[1]{\abinv} sample is assumed to be
fully in the $P(\Pem,\Pep) = (-80\%,+20\%)$ configuration.
\end{abstract}
%
% Introduction
%
\section{Introduction}
\label{sec:introduction}
The discovery of a Standard Model (SM)--like Higgs boson, announced on July 4th, 2012
by the ATLAS and CMS collaborations~\cite{Aad20121,Chatrchyan201230},
was celebrated as a major milestone in particle physics.
In the SM, the coupling strength of the Higgs boson to a fermion is given by
$y_{f}=\sqrt{2} m_{f}/v$, where $m_f$ is the fermion mass and
$v\approx \unit[246]{GeV}$ is the vacuum expectation value.
Since the top quark is the heaviest known elementary particle, the measurement of
the top Yukawa coupling, $\ytop$, serves as the high endpoint to test this prediction.
A sizable deviation in $\ytop$ from the SM prediction
is expected in various new physics scenarios,
which motivates a precise measurement of $\ytop$.
For example,
in composite Higgs models, where the Higgs boson is a pseudo-Nambu-Goldstone boson,
$\ytop$ could deviate up to tens of \%,
even in the scenario that no new particles are
discovered in LHC Run~2 data~\cite{PhysRevD.88.055024}.

A recent study of the prospects of measuring $\ytop$ at the LHC~\cite{CMS:2013xfa}
estimates that a precision of 14--15\% (7--10\%) is achievable
with an integrated luminosity of $\unit[0.3]{\abinv}$ ($\unit[3]{\abinv}$),
including theoretical and systematic uncertainties.
For $\epluseminus$ collisions,
detailed simulation studies have been carried out
using the $\ttH$ process at various center-of-mass energies.
At $\sqrt{s}=500$~GeV~\cite{Baer:1999ge,Juste:2006sv,Yonamine:2011jg},
where the $\epluseminus\rightarrow\ttH$ cross section is sharply rising,
the statistical precision is estimated to be about $10\%$
for an integrated luminosity of $\unit[1]{\abinv}$.
At $\sqrt{s}=800$~GeV~\cite{Juste:1999af,Gay:2006vs},
it is estimated that $\ytop$ can be measured to a precision of 5--6\%
for an integrated luminosity of $\unit[1]{\abinv}$,
including the systematic uncertainties due to the background normalization.

The International Linear Collider (ILC)~\cite{Adolphsen:2013kya} is a proposed $\epluseminus$ collider with a
maximum center-of-mass energy $\sqrt{s} = \unit[1]{TeV}$. It has a broad physics potential that is complementary to the LHC and
precision measurements of the Higgs couplings are an integral part of the physics program at this machine.
We present studies of the measurement of the top Yukawa coupling in direct observation at the \unit[1]{TeV} stage of the ILC.
The studies are carried out in ILD and SiD~\cite{Behnke:2013lya}, the two detector concepts for the ILC.
They are performed with detailed detector simulations taking into account the main beam-induced backgrounds at the collider
as well as the dominant background from other physics processes. Two final states are considered - events where both \PW bosons from the top quarks decay hadronically, and events where exactly one of the two \PW bosons decays leptonically.

The studies performed for the two detector concepts have large overlaps, and we highlight significant differences between the two analyses wherever applicable. This document is organized as follows: Section~\ref{sec:DataSamples} gives an overview over the signal sample and the considered physics background.
Section~\ref{sec:DetectorModel} gives brief overviews over the two ILC detector models. The tools for the generation of physics processes and the detector simulation and reconstruction are listed in Section~\ref{sec:AnalysisFramework}. The two dominant sources of machine-induced background in the detectors are introduced in Section~\ref{sec:beam_backgrounds}. The techniques to reduce these backgrounds and reconstruct the top quarks and Higgs bosons are described in Section~\ref{sec:EventReconstruction}. Details of the event selection are given in Section~\ref{sec:EventSelection} and the results are presented in Section~\ref{sec:Results}. The dominant sources of systematic uncertainty are given in Section~\ref{sec:SystematicUncertainties} and the two analyses are summarized in Section~\ref{sec:Summary}.
\section{Signal and Background Processes}
\label{sec:DataSamples}
\begin{figure}[htbp]
\centering
\includegraphics[width=0.45\linewidth]{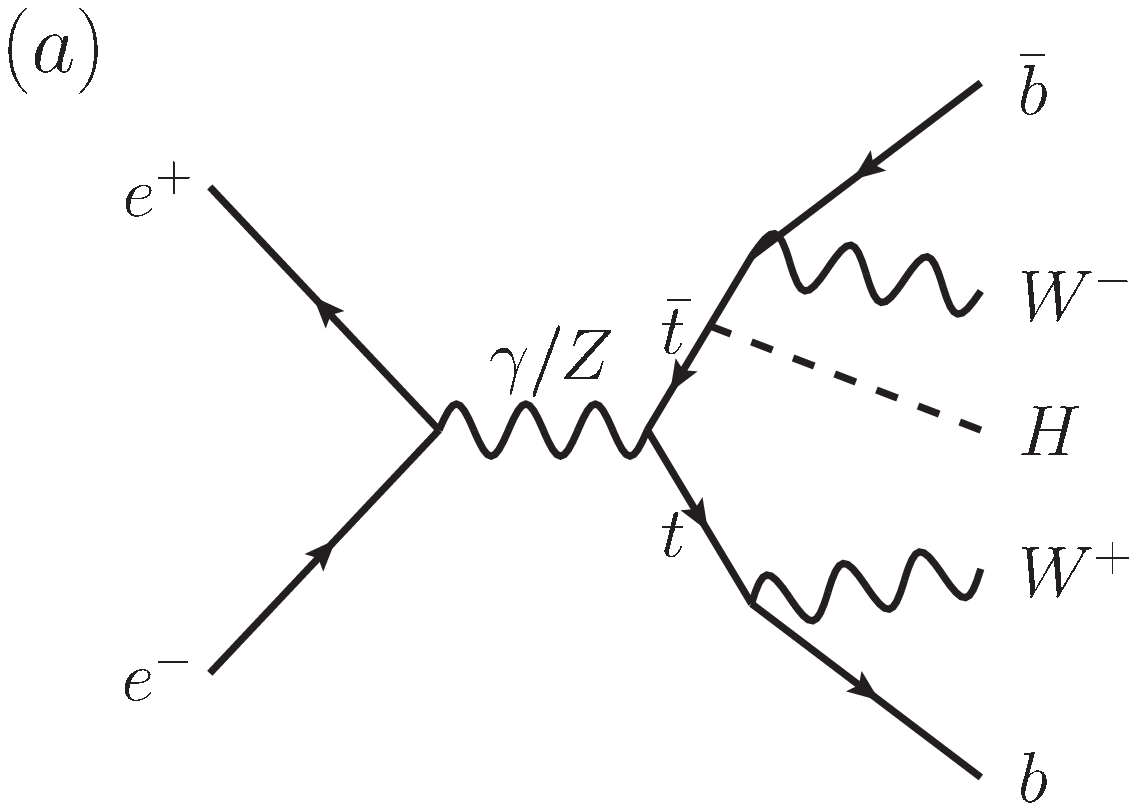}
\includegraphics[width=0.45\linewidth]{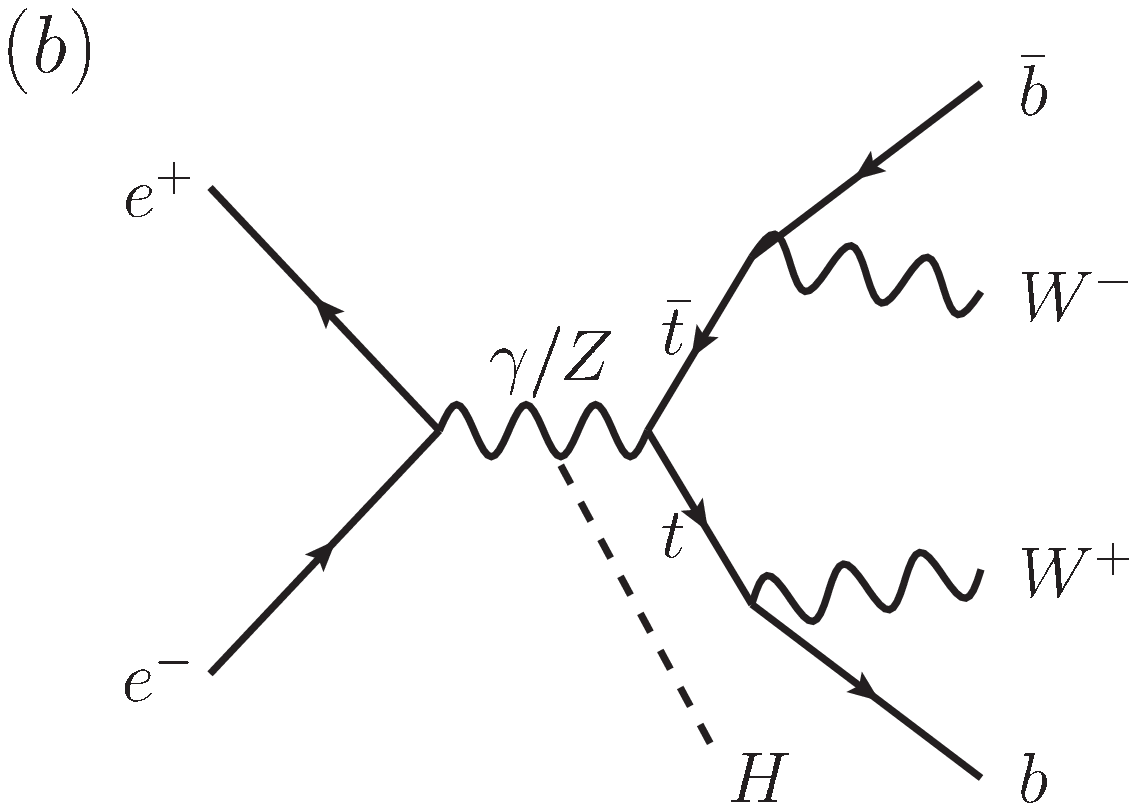}
\caption[The lowest order Feynman diagrams for the process $\ee\to\ttH$]{The lowest order Feynman diagrams for the process $\ee\to\ttH$. In (a) the Higgs boson is radiated from a top quark and (b) is the background Higgs-strahlung process where the Higgs boson is radiated from the Z boson.}
\label{fig:feyn_tth}
\end{figure}

Figure \ref{fig:feyn_tth} illustrates the lowest order Feynman diagrams for the process $\ee\to\ttH$.
The diagram for the Higgs-strahlung process $\ee\to\PZst\PH$ with $\PZst\to\ttbar$,
which does not depend on $\ytop$, has a small yet non-negligible
contribution to the total cross section.
The size of this effect is studied by evaluating the behavior of the $\ee\rightarrow\ttH$ cross section
when changing $\ytop$ from the SM value, using the linear approximation
${\Delta \ytop}/{\ytop}=\kappa\cdot{\Delta \sigma}/{\sigma}$.
In the absence of the Higgs-strahlung diagram, we would find $\kappa=0.5$.
Instead, we find $\kappa=0.52$, indicating a non-negligible contribution
from the Higgs-strahlung diagram to the total cross section at $\sqrt{s}=\unit[1]{TeV}$.
This factor is used in the extraction of the top Yukawa coupling precision.
The correction will be known with good precision,
because the Higgs coupling to the $\PZ$ boson can be extracted from measurements of
$\epluseminus \to \PZ\smH$ events at $\sqrt{s} = \unit[250]{GeV}$ with a statistical uncertainty of
about 1.5\%~\cite{Brau:2012hv}.

For this study the semi-leptonic and hadronic decays of the $\ttbar$ system
were studied with the Higgs decaying via the dominant decay mode into a $\bbbar$ pair.
For the fully hadronic decay channel this leads to a signature of eight hadronic jets, four of which are \PQb jets.
In the semi-leptonic mode the final signal in the detector consists of six hadronic jets, four of which are \PQb jets,
an isolated lepton, and missing energy and momentum from a neutrino.
For isolated leptons, only the prompt electrons and muons are reconstructed and considered as signal,
neglecting the decays into $\tau$ leptons.
These two modes are reconstructed in independent samples and are combined statistically.

Irreducible backgrounds to these processes arise from the eight-fermion final states of $\ttZ$ where the Z decays into a $\bbbar$ pair
and $\ttbar\bbbar$ where the $\ttbar$ system radiates a hard gluon which forms a $\bbbar$ pair.
A large background contribution also arises from $\ttbar$ due to the huge relative cross section compared to the signal.
There is also a contribution from the other decay modes of the $\ttH$ system such as the Higgs decaying to final states other than a $\bbbar$ pair and
the fully leptonic decays of the top quarks.

An overview of the cross sections (times branching ratio for the specifically listed final states) for the signal final states as well as for the
considered backgrounds is shown in Table~\ref{sid:benchmarking:tab:tth_cross_sections}.
For the measurement using the final state with six jets, all other $\ttH$ events, i.e., all
events where both top quarks decay either leptonically or hadronically, or events where the
Higgs boson does not decay into $\bpair$, are treated as background. For the eight-jets
final state events where at least one top quark decays leptonically or where the Higgs
boson decays into final states other than $\bpair$ are considered as background. The non-$\ttH$ backgrounds
are considered for both measurements.
\begin{table}
\centering
\caption{Production cross sections (times branching ratio for the specifically listed final states)
for the signal final states and the considered backgrounds. All samples were
generated assuming a Standard-Model Higgs boson with a mass of \unit[125]{GeV}. The numbers for ``other $\ttH$''
processes in this table do not include either of the signal final states (see text). The $\toppair \PZ$ and $\toppair \Pg^{*}$ samples,
where the hard gluon $\Pg^{*}$ splits into a $\bpair$ pair,
do not contain events where both top quarks decay leptonically. The $\toppair$ samples contain the SM
decays of both \PW bosons.}
\label{sid:benchmarking:tab:tth_cross_sections}
\begin{tabular}{l c c c c}
Type & Final state & P(\Pem) & P(\Pep) & $\sigma$ [$\times$ BR] (fb) \\ \hline
Signal & \ttH (8 jets) & $-80\%$ & $+20\%$ & 0.87 \\
Signal & \ttH (8 jets) & $+80\%$ & $-20\%$ & 0.44 \\
Signal & \ttH (6 jets) & $-80\%$ & $+20\%$ & 0.84 \\
Signal & \ttH (6 jets) & $+80\%$ & $-20\%$ & 0.42 \\ \hline
Background & other \ttH & $-80\%$ & $+20\%$ & 1.59 \\
Background & other \ttH & $+80\%$ & $-20\%$ & 0.80 \\
Background & $\toppair Z$ & $-80\%$ & $+20\%$ & 6.92 \\
Background & $\toppair Z$ & $+80\%$ & $-20\%$ & 2.61 \\
Background & $\toppair g^{*} \to \toppair\bpair$ & $-80\%$ & $+20\%$ & 1.72 \\
Background & $\toppair g^{*} \to \toppair\bpair$ & $+80\%$ & $-20\%$ & 0.86 \\
Background & $\toppair$ & $-80\%$ & $+20\%$ & 449 \\
Background & $\toppair$ & $+80\%$ & $-20\%$ & 170 \\
\end{tabular}
\end{table}

\section{Detector Models}
\label{sec:DetectorModel}
SiD~\citep[chapter 2]{Behnke:2013lya} and ILD concepts~\citep[chapter 3]{Behnke:2013lya} are
designed to be the two general-purpose detectors for the ILC,
with a \unit[4]{$\pi$} coverage, employing highly granular calorimeters for particle flow calorimetry.

For SiD
a superconducting solenoid with an inner radius of \unit[2.6]{m} provides a central magnetic field of \unit[5]{T}. The calorimeters are placed inside the coil and consist of a 30 layer tungsten--silicon electromagnetic calorimeter (ECAL) with \unit[13]{$\mathrm{mm}^2$} segmentation, followed by a hadronic calorimeter (HCAL) with steel absorber and instrumented with resistive plate chambers (RPC) -- 40 layers in the barrel region and 45 layers in the endcaps. The read-out cell size in the HCAL is \unit[$10\times10$]{$\mathrm{mm}^2$}. The iron return yoke outside of the coil is instrumented with 11 RPC layers with \unit[$30\times30$]{$\mathrm{mm}^2$} read-out cells for muon identification.
The silicon-only tracking system consists of five layers of \unit[$20\times20$]{$\micron^2$} pixels followed by five strip layers with a pitch of \unit[25]{\micron}, a read-out pitch of \unit[50]{\micron} and a length of \unit[92]{mm} per module in the barrel region. The tracking system in the endcap consists of four stereo-strip disks with similar pitch and a stereo angle of $12^\circ$, complemented by four pixelated disks in the vertex region with a pixel size of \unit[$20\times 20$]{$\micron^2$} and three disks in the far-forward region at lower radii with a pixel size of \unit[$50\times50$]{$\micron^2$}.
All sub-detectors have the capability of time-stamping at the level of individual bunches, \unit[337]{ns} apart, $\approx 1300$ to a train. This allows to separate hits originating from different bunch crossings. The whole detector will be read out in the \unit[200]{ms} between bunch trains.

The ILD detector model is designed around a different optimization with a larger size.
The ECAL and HCAL are placed inside
a superconducting solenoid, which provide a magnetic field of \unit[3.5]{T}.
The silicon-tungsten ECAL has an inner radius of \unit[1.8]{m} and a total thickness of \unit[20]{cm},
with $5\times\unit[5]{mm^2}$ transverse cell size and 30 layers of longitudinal segmentation.
The steel-scintillator HCAL has an outer radius of \unit[3.4]{m} with
$3\times\unit[3]{cm^2}$ transverse tiles and 48 layers longitudinal segmentation. % corresponding to 6 interaction length
ILD employs a hybrid tracking system consisting of a time projection chamber (TPC)
which provides up to 224 points per track and silicon-strip sensors for improved track momentum resolution,
which are placed in the barrel region both inside and outside the TPC
and in the endcap region outside the TPC.
The vertex detector consists of three double layers of silicon pixel sensors
with radii ranging from 15 to \unit[60]{mm}, providing a spatial resolution of \unit[2.8]{$\micron$}.
An iron return yoke instrumented with a muon detector and a tail catcher is placed outside the yoke.
In addition, silicon trackers and beam/luminosity calorimeters are installed in the forward region.

\section{Analysis Framework}
\label{sec:AnalysisFramework}
The $\ttH$, $\ttZ$, and $\ttbar\bbbar$ samples were generated using the \textsc{physsim}~\cite{physsim-homepage} event generator.
The sample referred to as $\ttbar$ in the following includes
six-fermion final states consistent with the $\ttbar$ decays
but not limited to the resonant $\ttbar$ production.
The $\ttbar$ events were generated using the \textsc{whizard 1.95}~\cite{Kilian:2007gr, Moretti:2001zz} event generator.
All samples were generated taking into account the expected beam energy spectrum at the $\sqrt{s}=\unit[1]{TeV}$ ILC, including initial state radiation~\cite{Skrzypek:1990qs} and beamstrahlung. The spectrum was sampled from a simulation of beam events~\cite{Schulte:1997tk}.
The model for the hadronization in \textsc{pythia 6.4}~\cite{Sjostrand:2006za} uses a tune based on \textsc{opal} data~\citep[Appendix B.3]{Linssen:2012hp}. % first described in a presentation, but never published ~\cite{Berggren:2010pt}

Detailed detector simulations based on \textsc{geant4}~\cite{Agostinelli:2002hh,Allison:2006ve} are performed.
In the SiD analysis, the event reconstruction is performed in the
\texttt{org.lcsim}~\cite{Graf:2011zzc} package.
The ILD analysis uses the \texttt{Marlin}~\cite{Gaede:2006pj,Aplin:2012kj} framework.
Both analyses use the \texttt{PandoraPFA}~\cite{Thomson:2009rp} algorithm for
calorimeter clustering and combined analysis of track and calorimeter information
based on the particle flow approach.
The \texttt{LCFIPlus} package~\citep[Section 2.2.2.3]{Behnke:2013lya}
is used for the tagging of heavy flavor jets.
The assumed integrated luminosity of the analysis is \unit[1]{\abinv},
which is split equally between the two polarization configurations
$(+80\%, -20\%)$ and $(-80\%, +20\%)$ for the polarization of the electron and positron beams $P(\Pem,\Pep)$.
Detector hits from Beam-induced backgrounds from processes described in Section~\ref{sec:beam_backgrounds} are treated correctly in the simulation of the detector readout and in the reconstruction.

\section{Simulation of Beam-Induced Backgrounds}
\label{sec:beam_backgrounds}
\begin{figure}
    \centering
    \begin{picture}(200,140)
    	\put(0,0){\includegraphics[width=.7\linewidth]{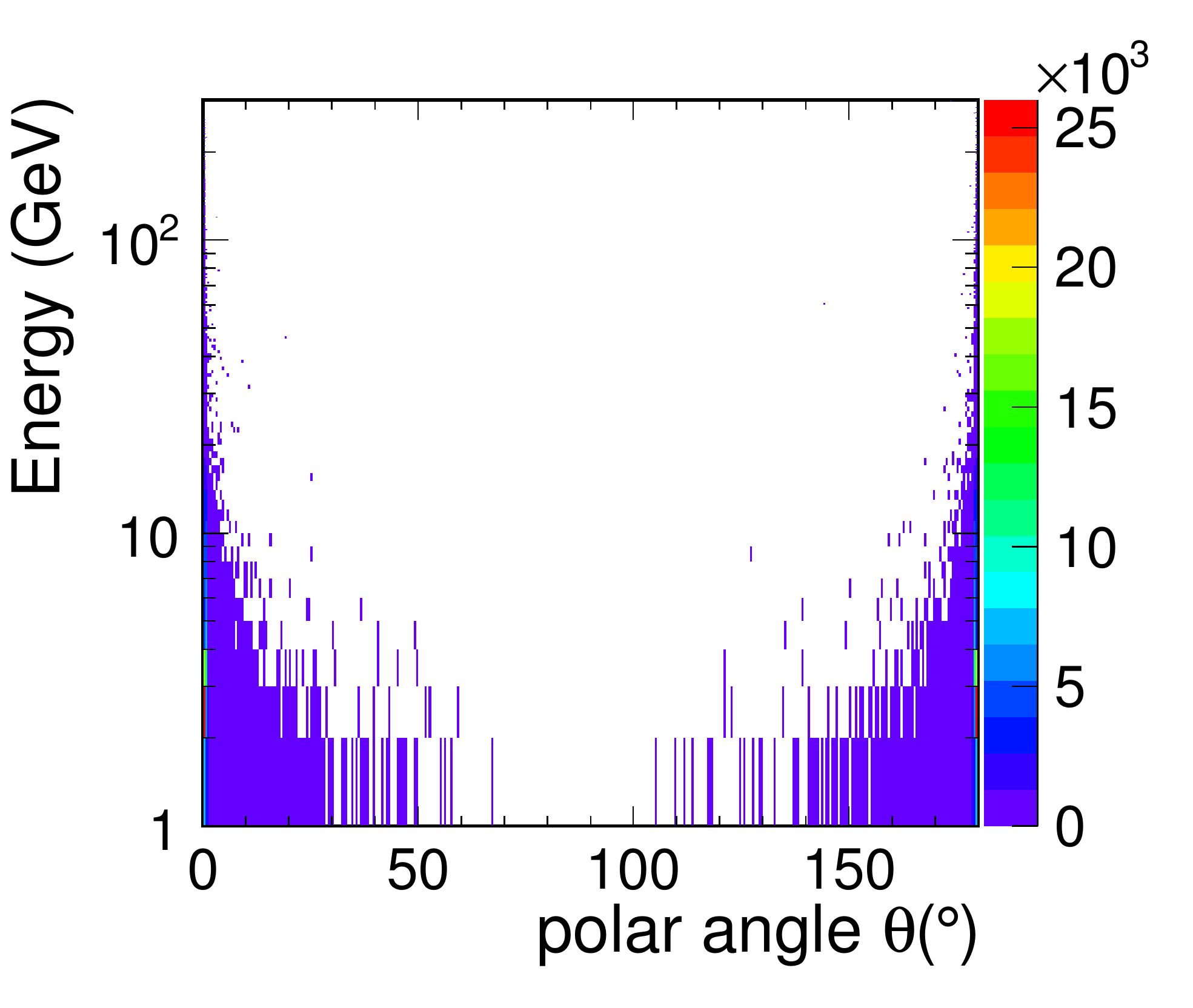}}
    	\put(40,112){\large(a)}
    \end{picture}
    \\
    \begin{picture}(200,140)
    	\put(0,0){\includegraphics[width=.7\linewidth]{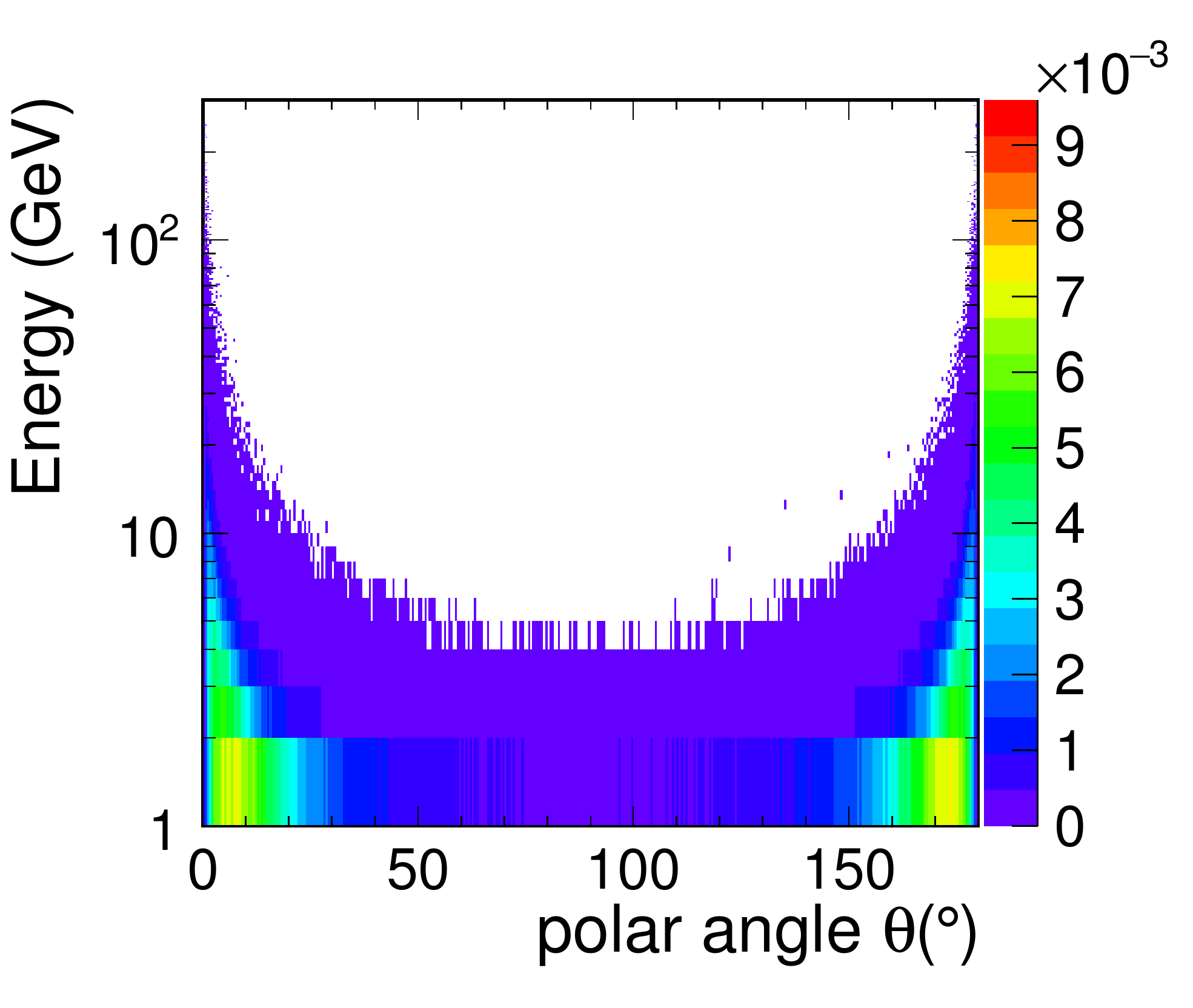}}
    	\put(40,112){\large(b)}
    \end{picture}
    \caption{The distributions of energy versus polar angle for the dominant beam-induced backgrounds. The contributions from incoherent \Pep\Pem production (a) and from \gghadrons processes (b) are shown separately. The number of entries in the histograms corresponds to the $4.5\times10^{5}$ (a) and 4.1 (b) events expected for one bunch crossing. The number of events in each bin in $(\theta, E)$ is shown in a color scale.}
    \label{fig:incoherentPairsGGHadMonteCarlo}
\end{figure}

The ILC operating at $\sqrt{s}=\unit[1]{TeV}$ has an instantaneous luminosity of $\unit[4.2 \times 10^{34}]{\text{cm}^{-2}\mathrm{s}^{-1}}$.
During the collision, a number of processes occur in addition to the primary scattering event.
The production of incoherent electron-positron pairs results in an average of $4.5\times 10^5$ low-momentum particles per bunch crossing.
We assume an average of 4.1 hadronic events from two-photon processes (\gghadrons) with a diphoton-invariant mass greater than \unit[300]{MeV}.
The distributions of the particles originating from these processes in the (polar angle, energy) plane are shown in Figure~\ref{fig:incoherentPairsGGHadMonteCarlo}.
They do not affect the reconstruction significantly, but present a challenge to the sub-detector occupancies and pattern recognition.
The SiD analysis includes both effects, while the ILD analysis includes only the \gghadrons processes.
The SiD analysis shows that incoherent pairs are under control by incorporating a detector design capable of time stamping individual bunch crossings.
%The possibility of using a detector design that integrates over multiple bunch crossings for the ILC is the subject of ongoing studies for ILD.
The baseline technology of the ILD barrel vertex detector integrates over 18 bunch crossings in layer 2 up to 180 bunch crossings in layers 3 -- 6.
Preliminary results of studies that take advantage of recent progress in the track reconstruction show a relative reduction of the b-tag purity at
80\% efficiency by about 7\%, while the c-tag purity at 60\% efficiency suffers a relative loss of 13\%~\cite{awlc14-strube-flavtag}.
Reducing the impact using advanced pattern recognition techniques in a detector design with smaller pixels that integrate over a
whole bunch train is the subject of ongoing efforts.

While the most energetic
particles from incoherent pair production are primarily outside of the detector acceptance of both detectors,
some low-$p_\mathrm{T}$ particles lead to an occupancy of up to \unit[0.06]{$\text{hits} / \mathrm{mm}^{2}$} per bunch crossing in the vertex detector and up to \unit[$5\times10^{-5}$]{$\text{hits}/\mathrm{mm}^{2}$} per bunch crossing in the main tracker for the SiD detector model. They do not, however, impact on the energy reconstruction. Particles from \gghadrons
processes on the other hand can have sizable values of $p_\mathrm{T}$ and reach the calorimeters,
affecting the jet energy resolution.
The beam-induced backgrounds do not degrade the tracking performance significantly~\cite{Behnke:2013lya}.

The primary vertices of the beam-induced backgrounds are distributed with a Gaussian profile along the beam direction across the luminous region of \unit[225]{\micron},
taking into account the bunch length along the beam direction.

\section{Event Reconstrucion}
\label{sec:EventReconstruction}
\subsection{Reconstruction of Isolated Leptons}
\label{sec:isolated_leptons}
Signal events with six jets contain one high-energy isolated lepton from the leptonic \PW boson decay.
No isolated leptons are expected in signal events with two hadronic \PW decays. Hence the number of isolated leptons is an important
observable in the signal selections for both final states.

The electron and muon identification criteria used in this study are based on the energy deposition
in the ECAL and HCAL and the momentum measured by the tracker.
Electron candidates are selected by requiring that almost all of the energy deposition is in the ECAL
and that the total calorimetric energy deposition is consistent with the momentum measured by the tracker.
For the muon candidates, most of the energy deposition is in the HCAL,
while the calorimetric energy is required to be small compared to the corresponding momentum measured by the tracker.
A selection on the impact parameter reduces non-prompt leptons.

The SiD analysis uses the \texttt{IsolatedLeptonFinder} processor implemented in
\texttt{MarlinReco}~\cite{Gaede:2006pj} to identify leptons in
regions with otherwise little calorimetric activity.
The ILD analysis additionally exploits the transverse distance from the jet axis to identify leptons from leptonic \PW decays.

The electron and muon identification capabilities of the reconstruction in a
multi-jet environment were tested in a sample of four jets, one lepton and
missing energy. The efficiency is defined as the fraction of
leptons with correctly identified flavor. The purity is defined as the ratio of the number of leptonic \PW
decays of a given flavor to the number of reconstructed isolated leptons
of that flavor. Leptons from heavy flavor meson decays and charged pions mis-identified as leptons are considered as background.
An efficiency of 82\% (89\%) and purity of 95\% (97\%) for electrons (muons) is observed in ILD
and 86\% (86\%) efficiency and 94\% (95\%) purity for electrons (muons) in SiD.

\subsection{Suppression of Beam-Induced Backgrounds}

\begin{figure}[htbp]
    \centering
    \includegraphics[width=0.5\textwidth]{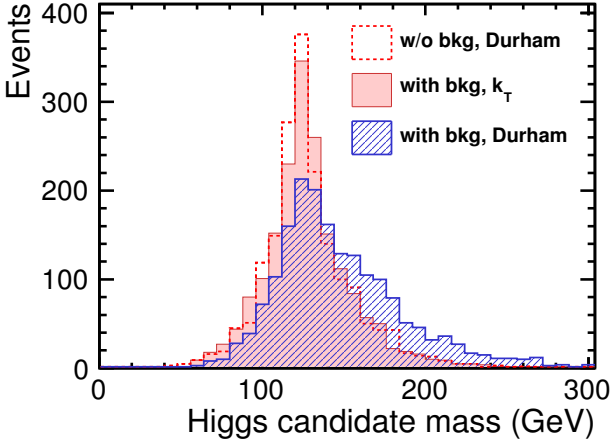}
    \caption{Distribution of the reconstructed Higgs mass in the ILD detector for signal events (six jets + lepton mode), without beam-induced backgrounds using the Durham algorithm, and with background, comparing the Durham and $k_T$ algorithms. By itself, the $k_T$ algorithm performs better than the Durham algorithm in terms of mitigating the effects of beam-induced backgrounds.}
    \label{fig:visenergygamgam}
\end{figure}

The particles from beam-induced backgrounds as described in Sec.~\ref{sec:beam_backgrounds} tend towards low transverse momenta and small angles with respect to the beam axis.
Different approaches are used to suppress the impact of the beam-induced backgrounds.
For the SiD analysis, only the reconstructed objects in the range $20^{\circ} < \theta < 160^{\circ}$ are considered,
because the {\ttH} final state is produced via $s$-channel exchange and is not suppressed by this selection.
In the ILD analysis,
the longitudinally-invariant $k_T$ jet algorithm~\cite{Catani:1993hr,Ellis:1993tq} with a value of 1.2 for the $R$ parameter is employed
to suppress the particles close to the beam axis. Only the particles grouped into the physics jets by the $k_T$ algorithm
are considered further in the analysis.
Figure~\ref{fig:visenergygamgam} shows how the impact of the beam-induced backgrounds on the reconstructed Higgs mass
is mitigated by the removal procedure.
A modified version of the Durham jet finding algorithm~\cite{Catani:1991hj} then groups all particles in the event into a specified number of jets, without splitting decay products of secondary vertices across different jets.
\subsection{Jet Clustering and Flavor Identification}
\label{sec:flavorIdentification}
Depending on the signal definition for the semi-leptonic or hadronic final state,
the Durham jet clustering algorithm is used in the exclusive mode to
cluster the event into six or eight jets, respectively. In either case, the isolated leptons described in Section~\ref{sec:isolated_leptons} are removed before the jet finding steps.

Heavy flavor identification is primarily used to remove the $\ttbar$ background.
Both the six-jet and eight-jet final states contain four \PQb jets.
The $\ttbar$ events contain no more than two \PQb jets from the top decays as do $\sim$80\% of $\ttZ$.
The flavor tagging classifier for the measurement of \ttH production was trained on events with six quarks of the same flavor produced in electron-positron annihilation. For the training, 60000 \PQc and \PQb jets, and 180000 light flavor quark jets are used. These samples were chosen since the jets have similar kinematic properties as those in \ttH signal events.
For a b-jet identification efficiency of 50\%,
the misidentification fraction is found to be about
0.12\% for c-jets and about 0.05\% for light quark (uds) jets for both detectors,
evaluated using two-jet final states at $\sqrt{s}=\unit[91]{GeV}$~\cite{Behnke:2013lya}.
The misidentification fraction for c-jets (uds-jets) increases by 100\% (60\%) when
incoherent electron-positron pairs and \gghadrons processes are included in the simulation as
described in Section~\ref{sec:beam_backgrounds}.
For the six-quark final states at $\sqrt{s}=\unit[1]{TeV}$, the misidentification fractions
typically degrade by a factor of two in addition
due to the varying jet energies and
the confusion in the jet clustering
due to the increased number of jets.

Figure~\ref{fig:cutvars}(a) shows the distribution of the response from the flavor-tagging multivariate selection for the jet that has the third-highest tagging probability. In both analyses, the shape of the distribution of the flavor tagging response, rather than a simple cut, is used.
The background channels, in particular \ttbar, are dominated by the peak at low values. The peak at higher values in the \ttZ channel
is due to events with four genuine \PQb jets.

\subsection{Reconstruction of W, top and Higgs Candidates}
\label{sec:reco_w_top_higgs}
To form \PW, top and Higgs candidates from the reconstructed jets, the following
function is minimized for the final state with eight jets:
\begin{align}
\chi^{2}_\text{8 jets} & = \frac{(M_{12}-M_{\PW})^{2}}{\sigma_{\PW}^{2}} + \frac{(M_{123}-M_{\PQt})^{2}}{\sigma_{\PQt}^{2}} + \frac{(M_{45}-M_{\PW})^{2}}{\sigma_{\PW}^{2}}\nonumber\\ & + \frac{(M_{456}-M_{\PQt})^{2}}{\sigma_{\PQt}^{2}} + \frac{(M_{78}-M_{\smH})^{2}}{\sigma_{\smH}^{2}},
\label{sid:benchmarking:eq:tth_eight_jet_pairing}
\end{align}
where $M_{12}$ and $M_{45}$ are the invariant masses of the jet pairs used to
reconstructed the \PW candidates, $M_{123}$ and $M_{456}$ are the invariant
masses of the three jets used to reconstruct the top candidates and $M_{78}$ is
the invariant mass of the jet pair used to reconstruct the Higgs candidate.
$M_{\PW}$, $M_{\PQt}$ and $M_{\smH}$ are the nominal \PW, top and Higgs masses.
The resolutions $\sigma_{\PW}$, $\sigma_{\PQt}$ and $\sigma_{\smH}$ were
obtained from reconstructed jet combinations matched to \PW, top and Higgs particles at
generator level. The corresponding function minimized for the six-jets final state
is given by:
\begin{equation}
\chi^{2}_{\textnormal{6 jets}} = \frac{(M_{12}-M_{\PW})^{2}}{\sigma_{\PW}^{2}} + \frac{(M_{123}-M_{\PQt})^{2}}{\sigma_{\PQt}^{2}} + \frac{(M_{45}-M_{\smH})^{2}}{\sigma_{\smH}^{2}}.
\label{sid:benchmarking:eq:tth_six_jet_pairing}
\end{equation}
In the ILD analysis, the $\PQb$ tagging information is also used to reduce the number of combinations by forming the Higgs candidate from two of the four jets with the highest value of the \PQb-tagging classifier. The other jets in the event are used to form the hadronic top candidates.

\section{Event Selection}
\label{sec:EventSelection}
\begin{figure}[htbp]
\centering
    \begin{picture}(100,100)
      \put(0,0){\includegraphics[width=.49\columnwidth]{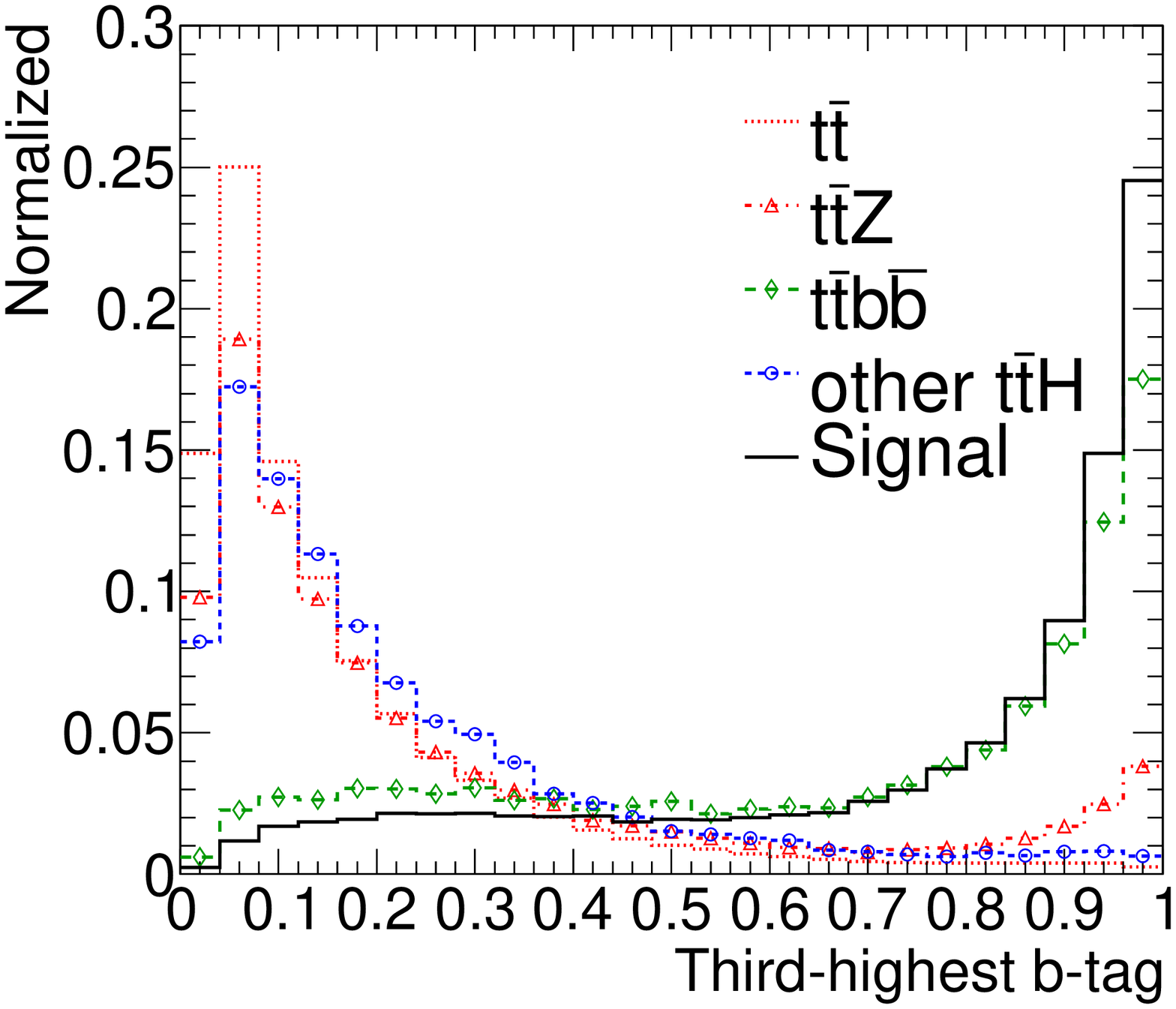}}
      \put(30,80){\large(a)}
    \end{picture}
    \hfill
    \begin{picture}(100,100)
      \put(0,0){\includegraphics[width=.49\columnwidth]{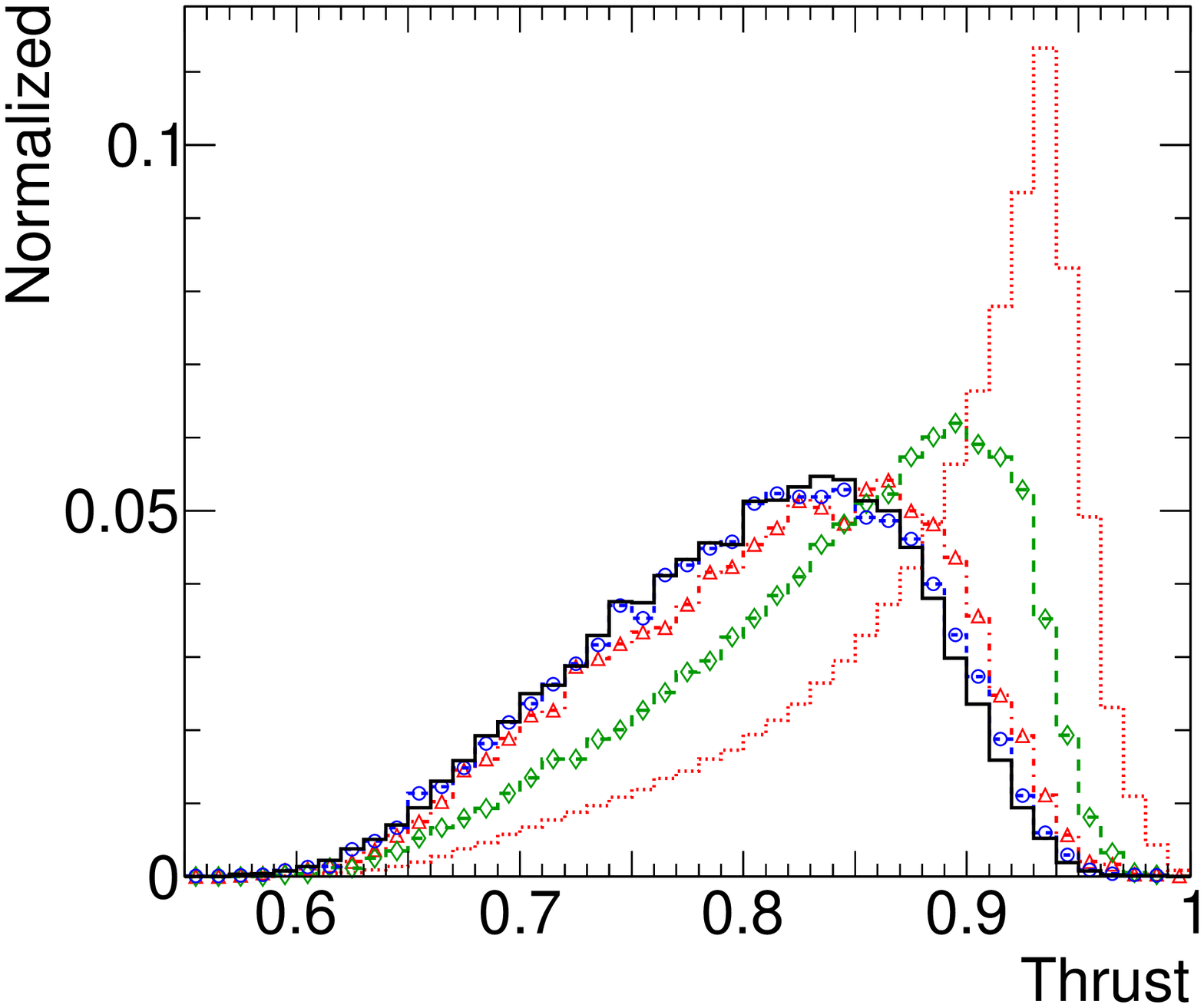}}
      \put(30,80){\large(b)}
    \end{picture} \\
    \begin{picture}(100,100)
      \put(0,0){\includegraphics[width=.49\columnwidth]{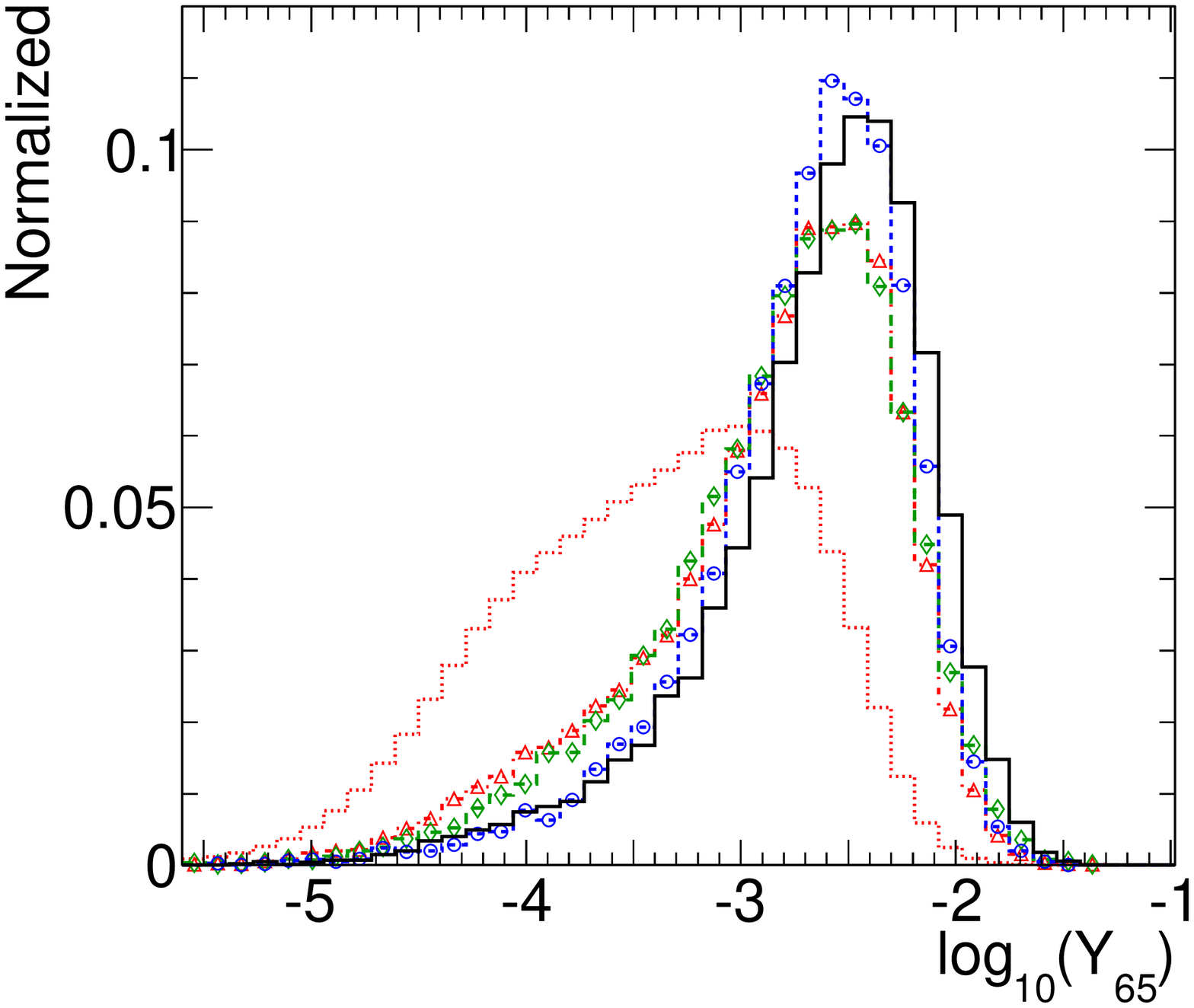}}
      \put(30,80){\large(c)}
    \end{picture}
    \hfill
    \begin{picture}(100,100)
      \put(0,0){\includegraphics[width=.49\columnwidth]{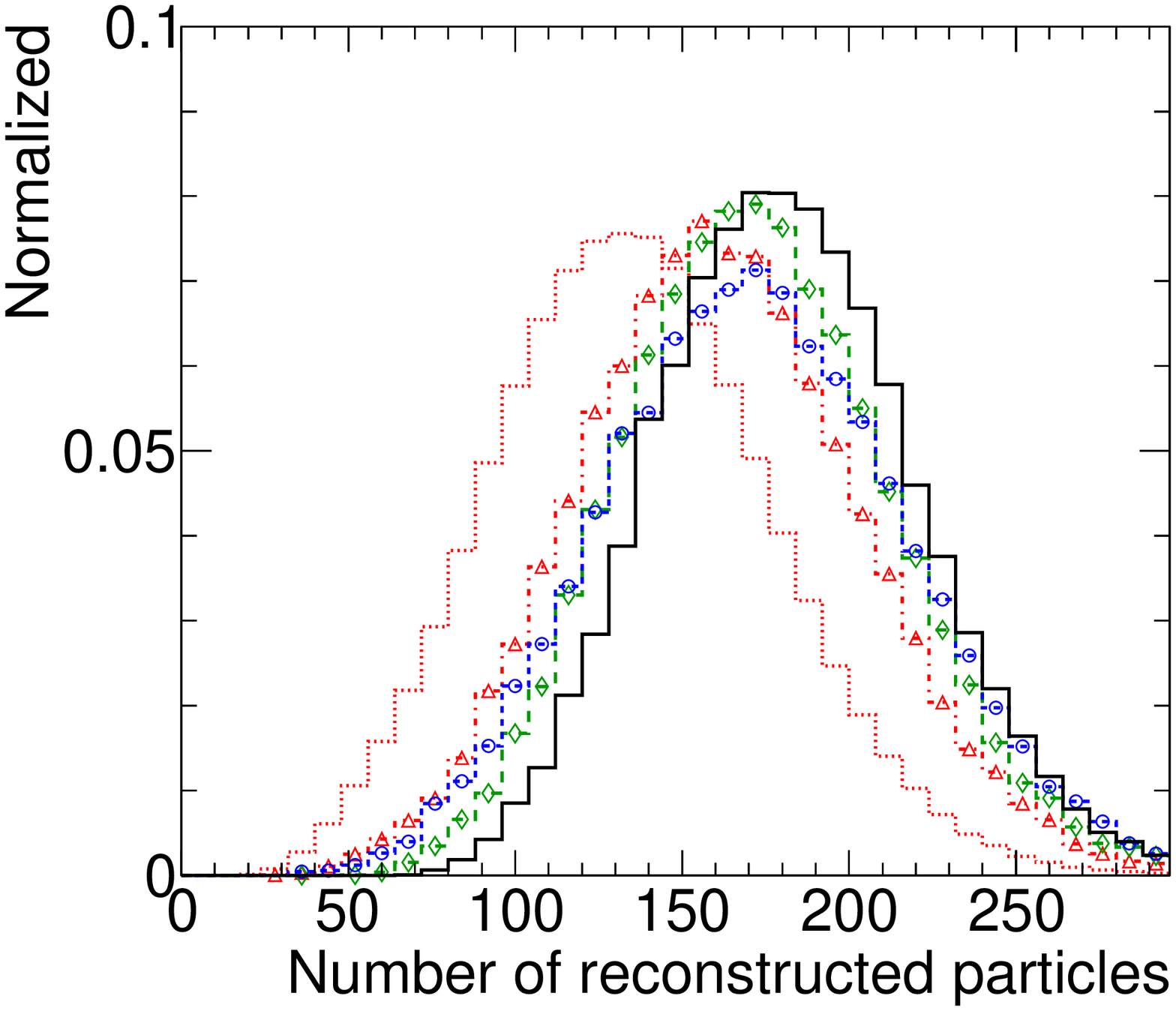}}
      \put(30,80){\large(d)}
    \end{picture}
\caption{Distributions of the event selection variables for the different signal and background processes in the ILD detector:
(a) the third highest b-tag in the event; (b) the event thrust; (c) the jet resolution parameter $Y_{65}$; (d) the number of reconstructed particles in the event.  All histograms are normalized to unit area.}
\label{fig:cutvars}
\end{figure}

Events were selected using Boosted Decision Trees (BDTs) as implemented in TMVA~\cite{Hocker:2007ht}.
The BDTs were trained separately for the
eight- and six-jets final states. The following input variables were used:
\begin{itemize}
\item the four highest values of the \PQb-tagging classifier. The third (see Figure~\ref{fig:cutvars}(a)) and fourth highest \PQb-tag values are especially suited to reject $\toppair$ and most of the $\ttZ$ events, both of which contain only two \PQb jets;
\item the event thrust~\cite{1964PhL....12...57B} (see Figure~\ref{fig:cutvars}(b)) defined as
\begin{equation}
    T = \max\frac{\sum_{i}|\hat{n}\cdot\vec{p_{i}}|}{\sum_{i}|\vec{p_{i}}|},
\end{equation}
where $p_{i}$ is the momentum of the jet. Since the top quarks in $\toppair$ events are produced back-to-back, the thrust variable has larger values in $\toppair$ events compared to $\ttH$, $\toppair Z$ or $\toppair \bpair$ events;
\item the jet resolution parameter from the Durham algorithm in the E recombination scheme $Y_{ij}$, when combining $i$ jets to $j = (i - 1)$ jets. For the six-jets final state $Y_{54}$ and $Y_{65}$ (see Figure~\ref{fig:cutvars}(c)) are found to be effective, while $Y_{76}$ and $Y_{87}$ are used for the eight-jets final state. Isolated leptons are removed prior to the jet clustering;
\item the number of identified isolated electrons and muons (ILD only);
\item the missing transverse momentum, $p_\mathrm{T}^{\textnormal{miss}}$. Due to the leptonic \PW boson decay, finite values of $p_\mathrm{T}^{\textnormal{miss}}$ are reconstructed for six-jets signal events while $p_\mathrm{T}^{\textnormal{miss}}$ tends towards zero for eight-jets signal events;
\item the visible energy of the event defined as the scalar sum of all jet energies;
\item the masses $M_{12}$, $M_{123}$ and $M_{45}$ as defined in Section~\ref{sec:reco_w_top_higgs}.
\end{itemize}
For the eight-jets final state additionally the two variables $M_{456}$ and $M_{78}$ as defined in Section~\ref{sec:reco_w_top_higgs} are included.

The ILD analysis includes the helicity angle of the Higgs candidate
as defined by the angle between the two \PQb jet momenta in the dijet rest frame.

\begin{figure}[htbp]
\centering
    \begin{picture}(100,100)
      \put(0,0){\includegraphics[width=.49\columnwidth]{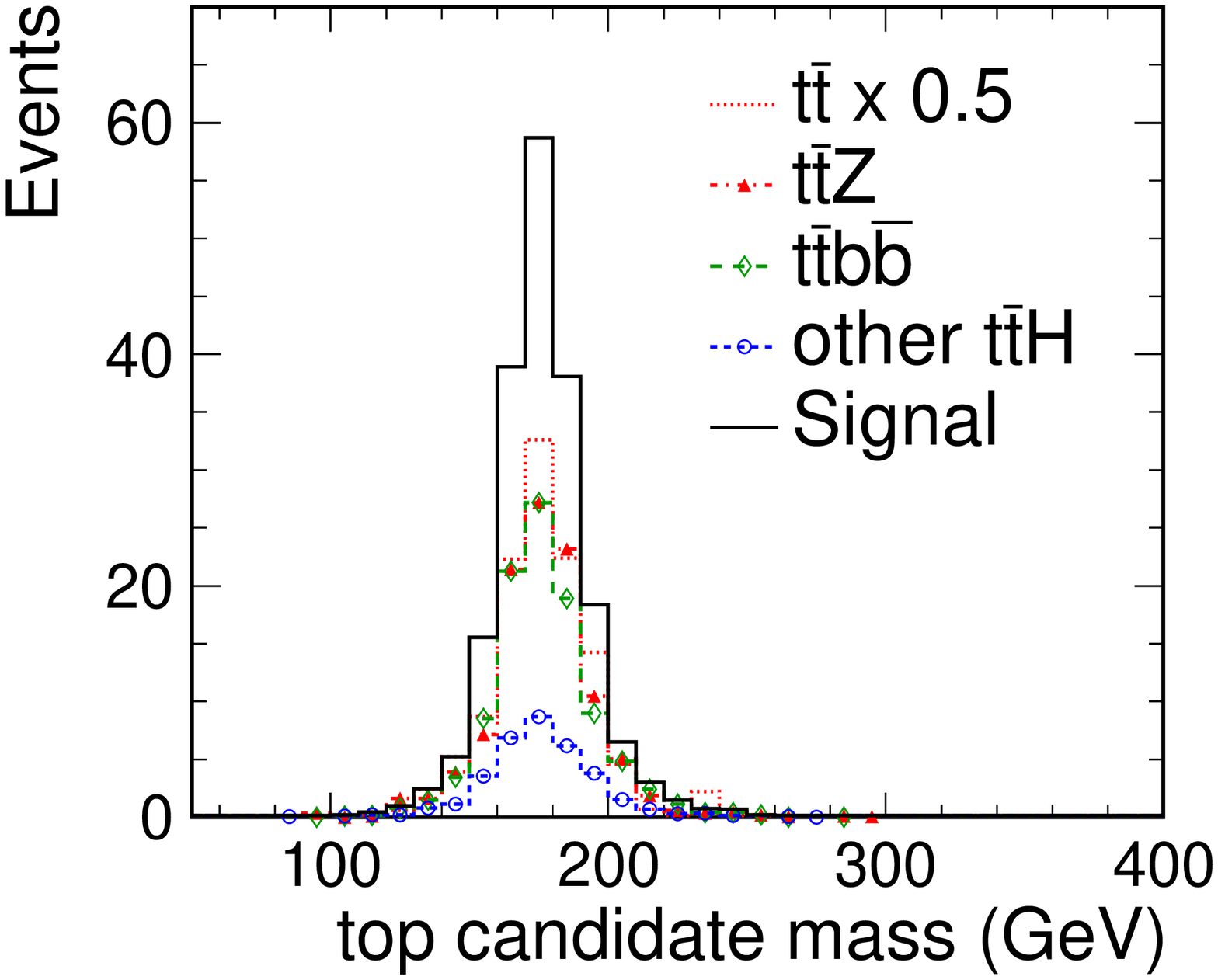}}
      \put(30,80){\large(a)}
    \end{picture}
    \hfill
    \begin{picture}(100,100)
      \put(0,0){\includegraphics[width=.49\columnwidth]{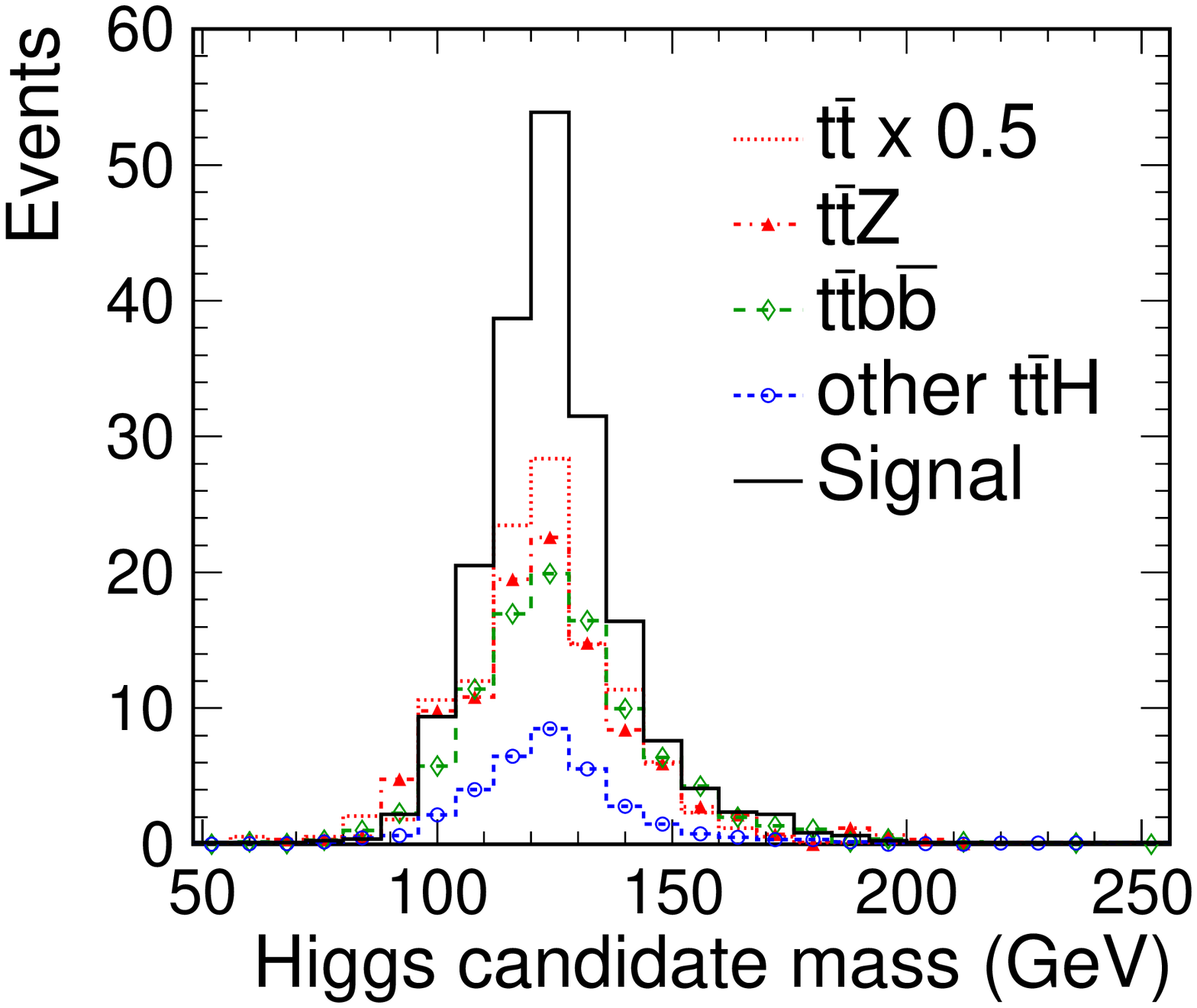}}
      \put(30,80){\large(b)}
    \end{picture}
\caption{\label{sid:benchmarking:fig:bt_outputs_tth_sixJet_mass} Reconstructed top (a) and Higgs (b) masses for selected six-jets events in the SiD detector. All histograms were normalized to an integrated luminosity of 1~\abinv. The distribution for \toppair was scaled by a factor of $0.5$.}
\end{figure}

To select events, cuts on the BDT response are applied. The cuts were
optimized by maximizing the signal significance given by:
$\frac{S}{\sqrt{S+B}}$, where $S$ is the number of signal events and $B$ is the
number of background events. As an example, the reconstructed top and Higgs masses in
six-jets events after the cut on the BDT output are shown in Figure~\ref{sid:benchmarking:fig:bt_outputs_tth_sixJet_mass}.
The selection efficiencies (purities) for signal events are 33.1\% (27.7\%) and 56.0\% (25.2\%) for the six- and eight-jets analyses in ILD, respectively, and 30.5\% (28.9\%) and 45.9\% (26.7\%) in SiD.
 In Table~\ref{tab:tth_selected_events} the expected yields are shown separately for all
investigated final states.
\begin{table}[htbp]
\centering
\caption{Number of selected events for the different
final states assuming an integrated luminosity of \unit[1]{\abinv}. The values obtained for the six-
and eight-jets final state selections are shown separately.}
\label{tab:tth_selected_events}
\begin{tabular}{l.|..|..}
\multicolumn{2}{l}{Detector}     & \multicolumn{2}{c}{ILD} & \multicolumn{2}{c}{SiD} \\
\hline
Sample                & \multicolumn{1}{c|}{\text{Before cuts}}      & \multicolumn{4}{c}{\text{After Cuts}} \\
                &     & \multicolumn{1}{c}{\text{6 jets}} &  \multicolumn{1}{c|}{\text{8 jets}} & \multicolumn{1}{c}{\text{6 jets}} & \multicolumn{1}{c}{\text{8 jets}} \\
\hline
$\ttH$ 6 jets         & 628.7    & 208.0 & 65.5  & 191.6 & 57.4 \\
$\ttH$ 8 jets         & 652.7    & 2.1   & 365.6 & 1.6   & 299.4 \\
$\ttH\to\text{other}$ & 1197.5   & 28.8  & 25.3  & 33.0 & 16.6 \\
$\ttZ$                & 5332.4   & 126.1 & 260.5 & 105.6 & 187.1 \\
$\ttbar\bbbar$        & 1434.5   & 125.4 & 222.6 & 100.1 & 180.7 \\
$\ttbar$              & 308800.9 & 261.2 & 513.6 & 232.0 & 381.6 \\
\hline
\multicolumn{2}{l|}{$\ytop$ statistical uncertainty} & \multicolumn{1}{r}{6.9\%} & \multicolumn{1}{r|}{5.4\%} & \multicolumn{1}{r}{7.0\%} & \multicolumn{1}{r}{5.8\%} \\
combined              &   & \multicolumn{2}{c|}{4.3\%} & \multicolumn{2}{c}{4.5\%} \\
\end{tabular}
\end{table}

\section{Results}
\label{sec:Results}
The cross section can be directly obtained from the number of
background-subtracted signal events after the selection. The
uncertainty of the cross section measurement was estimated using the number
of selected signal and background events. Assuming an
integrated luminosity of 1~\abinv split equally between the \mbox{$P(\Pem,\Pep) = (-80\%,+20\%)$}
and \mbox{$P(\Pem,\Pep) = (+80\%,-20\%)$} beam polarization configurations, the cross section can be measured with a
statistical precision of 10 -- 11\% using the eight-jets final state and with a
statistical precision of $\approx 13\%$ for the six-jets final state.

The uncertainties of the measured cross sections translate to
precisions on the top Yukawa coupling of 5 -- 6\% and $\approx 7\%$ from the eight- and
six-jets final states, respectively. If both measurements are combined, the top
Yukawa coupling can be extracted with a statistical precision of better than $4.5\%$.

For 1~\abinv of data with only $P(\Pem,\Pep) = (-80\%,+20\%)$ polarization, this
number improves to $4\%$.
The precision for the six-jets final state could be improved further if
$\tau$ leptons were included in the reconstruction. Additional improvement
is also foreseen by employing kinematic fitting.
%The achieved precision of both analyses indicates that the reconstruction of the investigated final states is limited by confusion when combining the jets to form top, Higgs and $\PW$ boson candidates rather than the differences in subdetector performance~\cite{Behnke:2013lya} between the two investigated concepts.
The achieved precision of both analyses indicates that the reconstruction
of the investigated final states is not significantly affected by the
differences in subdetector performance~\cite{Behnke:2013lya} between the
two investigated concepts. This is consistent with the findings of the
study of the Higgs self-coupling in the six-jets final state of the
$\PZ\PH\PH$ channel~\cite[Chapter 2.5.2]{Baer:2013cma}, where the
confusion in the jet clustering was a dominant contribution to the
mass resolution.

\section{Systematic Uncertainties}
\label{sec:SystematicUncertainties}
Given the low cross section and relatively clean environment at a $\sqrt{s}=\unit[1]{TeV}$ ILC,
it is expected that the statistical uncertainty of the measurement of the top Yukawa coupling
in direct observation dominates over systematic uncertainty.
In the following we estimate the contributions from the main sources of systematic errors to this measurement.

The number of background events in the final selection is comparable to the number of
signal events, making the estimation of normalization and shapes of the background an important source of systematic uncertainty.
%
%The total cross section is expected to be calculable from theory
%to very good precision for the $\ttZ$ and $\ttbar$ processes.
The total cross sections calculable from theory
for the $\ttZ$ and $\ttbar$ processes
are expected to improve in the coming years.
For the $\ttZ$ process,
the QCD and electroweak corrections are known at the 1-loop level~\cite{Lei:2008jx}.
%+5.0\% and -5.4\%, respectively, at $\sqrt{s}=\unit[1.2]{TeV}$ for a Higgs mass of \unit[120]{GeV}.
%
For the $\ttbar$ process,
the electroweak corrections are known to the 1-loop level~\cite{Beenakker:1991ca},
while the QCD corrections are known at the 3-loop level~\cite{Grunberg:1979ru,Jersak:1981sp,Chetyrkin:1996cf,Chetyrkin:1997mb,Chetyrkin:1997pn}.
QCD contributions to the $\ttbar\bbbar$ cross section make this value more challenging to compute precisely;
in principle the measurement of the gluon splitting rate
at relevant energies will provide a handle to estimate its size.
A crucial aspect in the estimation of the efficiencies is the accurate modeling of the event selection variables.
Here we illustrate how one might arrive at control samples for different background sources
in order to estimate the efficiency of each component accurately.

The \ttZ final state can be reconstructed in a similar fashion to the $\ttH$ final state.
For hadronically decaying \PZ, the number of jets in the final state will be the same as in the $\ttH$ analysis.
For our nominal integrated luminosities of \unit[0.5]{\abinv} for each of the two polarization states,
nearly 1300 signal events $\ttH(\rightarrow\bbbar)$ are expected
and 800 events are expected for $\ttZ(\rightarrow\bbbar)$,
taking into account the $\PZ\rightarrow\bbbar$ branching ratio.
Other hadronic decays of the \PZ boson will have large $\ttbar$ background
due to the absence of the two \PQb tags. Including leptonic decays of the
\PZ boson will help increase the sensitivity to this channel.
Overall, one can expect that the statistical uncertainty for \ttZ will be similar to that of \ttH,
i.e. at the few percent level.

The large cross section of $\ttbar$ events will allow for detailed systematic studies.
While only a certain class of these events may enter the final selection, we estimate
that the systematic uncertainty to the measurement of the top Yukawa coupling
is comparable to that of $\ttZ$.

Other sources of systematic uncertainty such as
the luminosity measurement, jet energy scale, and flavor tagging
are typically at the 1\% level or better for $\ee$ colliders~\cite{ALEPH:2005ab}.
The uncertainty on BR($\smH \to \bpair$) is not taken into account in our calculation of the top Yukawa coupling
from the \ttH production cross section. It is expected that this quantity
can be measured with a precision of better than $1\%$ using $\epluseminus \to \PGn\PAGn\smH$
events~\cite{Barklow:2003hz, lc-rep-2013-005}.

\section{Summary}
\label{sec:Summary}
The physics potential for a measurement of the top Yukawa coupling at \unit[1]{TeV} at the ILC is investigated. The study is based on detailed detector simulations using both the SiD and ILD detector concepts. Beam-induced backgrounds are considered in the analysis. The combination of results obtained for two different final states leads to a statistical uncertainty on the top Yukawa coupling of better than 4.5\% for an integrated luminosity of 0.5~\abinv with the $P(\Pem,\Pep) = (-80\%,+20\%)$ beam polarization configuration and 0.5~\abinv with $P(\Pem,\Pep) = (+80\%,-20\%)$ polarization. If 1~\abinv of data were recorded with only the $P(\Pem,\Pep) = (-80\%,+20\%)$ beam polarization configuration, the expected precision would improve to $4\%$.

The results from the studies presented in this paper demonstrate the robustness of the physics reconstruction of high jet multiplicity final states at $\sqrt{s}=1$~TeV under realistic simulation conditions. The expected precisions for measurements of the top Yukawa coupling were found to be very similar for two different detector concepts.
\begin{acknowledgements}
The authors would like to thank their colleagues in the
Linear Collider community for their help in facilitating this work;
in particular, T.~Barklow, M.~Berggren, and A.~Miyamoto for generating the Monte-Carlo samples;
J.~Engels, C.~Grefe, and S.~Poss for the production on the Grid.
The authors also thank K.~Fujii, N.~Watson, and V.~Martin for the helpful discussions and suggestions.
This work was partially supported by JSPS KAKENHI Grant Number 23000002.
\end{acknowledgements}
%
% References
%
\raggedleft
\bibliographystyle{spphys}
\bibliography{tth_dbd_paper}
\end{document}